\documentstyle[12pt,psfig]{article}
\newcommand{\beq}{\begin{equation}}
\newcommand{\eeq}{\end{equation}}
\newcommand{\beqa}{\begin{eqnarray}}
\newcommand{\eeqa}{\end{eqnarray}}
\newcommand{\ba}{\begin{array}}
\newcommand{\ea}{\end{array}}
\newcommand{\CR}{\nonumber \\}

\renewcommand{\thefootnote}{\fnsymbol{footnote}}

\makeatletter
\@addtoreset{equation}{section}

\makeatother

\begin{document}

\begin{titlepage}
\null
\begin{flushright}
YITP-00-61
\\
UT-915
\\
hep-th/0011101
\\
November, 2000
\end{flushright}

\vskip 1.5cm
\begin{center}


  {\LARGE Landau-Ginzburg Description of D-branes \medskip\\ on ALE Spaces}

\lineskip .75em
\vskip 2.5cm
\normalsize

  {\large Jiro Hashiba$^1$ and Michihiro Naka$^2$}

\vskip 1.5em

  $^1${Yukawa Institute for Theoretical Physics,\\
               Kyoto University, Kyoto 606-8502, Japan\\
           {\normalsize hashiba@yukawa.kyoto-u.ac.jp}}
\vskip 1.5em

  $^2${Department of Physics, University of Tokyo\\
               Tokyo 113-0033, Japan\\
              {naka@hep-th.phys.s.u-tokyo.ac.jp}}

\vskip 2cm

{\bf Abstract}

\end{center}

We study the Landau-Ginzburg (LG) mirror theory of the non-linear sigma
model on the ALE space ${\cal M}$ obtained by resolving the singularity
of the orbifold ${\bf C}^2/{\bf Z}_N$. In the LG description, the data
of the BPS spectrum and the lines of marginal stability are encoded in the
special Lagrangian submanifolds of the mirror manifold $\widehat{{\cal M}}$. 
Our LG description is quite simple as compared with 
the quiver gauge theory analysis near the orbifold point.
Furthermore our mirror analysis can be applied to any point 
on the moduli space of ${\cal M}$. 
\end{titlepage}

\renewcommand{\thefootnote}{\arabic{footnote}}
\baselineskip=0.6cm

\clearpage

\section{Introduction}
\hspace{5mm}
In the Calabi-Yau (CY) compactifications of string theory, 
a deep understanding of D-branes wrapped around the compact submanifolds 
is required to determine the physical BPS spectrum. 
When one moves around the compactification moduli space,
the nature of wrapped D-branes is expected to change dramatically. 
On the moduli space, there are two distinguished points which
have been physically well understood: ${\it large~volume~limit}$ and
${\it orbifold~point}$.
One may have a natural question how
one should interpolate these two points 
to characterize BPS D-branes at arbitrary points on the moduli space.
Recently, many analyses concerning this problem have been performed in
\cite{BrDoLaRo, DiGo, DiRo, DoFiRo, DoFiRo2, De, MoOhYa}. 

In these analyses, mirror symmetry has played a central role. 
One of the recent developments in this direction is found in 
\cite{HoVa,HoIqVa}, where the non-linear sigma model on a toric variety
is shown to have an equivalent description given by a Landau-Ginzburg
(LG) model with a suitable superpotential.
Furthermore, it has been realized there that 
D-branes wrapping holomorphic cycles in the toric variety are mirror to 
D-branes wrapped on certain Lagrangian submanifolds on the LG side.
The works \cite{GoJaSa} treat a related subject.

In this paper, we wish to utilize this mirror technique to study the BPS 
spectrum coming from D-branes on the ALE space ${\cal M}$ given by
blowing up the singularity of the orbifold ${\bf C}^2/{\bf Z}_N$. 
In \cite{FiMa}, D-branes on ${\cal M}$ near the orbifold point have been
investigated in terms of the quiver gauge theory \cite{DoMo}.
Moreover the BPS spectrum and the lines of marginal stability 
were determined.
In particular, it has been recognized that all the lines do not lead to
the decay of BPS states.
However, the analysis has been restricted to the vicinity
of the orbifold point in
the whole moduli space spanned by complex blowing up parameters.
The quiver gauge theory inevitably fixes the value 
of $B$ field in the non-linear sigma model on ${\cal M}$.

A distinguished feature of using the LG mirror description is that
we can examine the BPS spectrum over the ${\it whole}$ moduli space spanned
by the blowing up parameters. 
We map the data of the BPS spectrum into the mirror manifold 
${\widehat{\cal M}}$.
Then, we give the mass formula and the lines of marginal stability 
in terms of the geometry of the special Lagrangian submanifolds 
in $\widehat{{\cal M}}$.

This paper is organized as follows. 
In section 2, we formulate the non-linear sigma model 
on the ALE space ${\cal M}$ using a certain gauged linear sigma model.
In section 3, 
we will explain how BPS spectrum near the orbifold point are 
studied by the representation of a quiver diagram, following \cite{FiMa}.
In section 4, the LG mirror description
of the sigma model on ${\cal M}$ will be presented, following
\cite{HoVa,HoIqVa}. 
We study BPS spectrum in terms of the special Lagrangian submanifolds in 
$\widehat{\cal M}$.
Then, we perform explicit analyses for the simplest cases, 
${\bf C}^2/{\bf Z}_2$ and ${\bf C}^2/{\bf Z}_3$. 
The last section is devoted to conclusions and discussion.

\section{Non-linear sigma model}
\hspace{5mm}
The non-linear sigma model on the ALE space ${\cal M}$, which is given
by blowing up the orbifold ${\bf C}^2/{\bf Z}_N$, is described by the
low energy effective theory of a certain ${\cal N}=2$
gauged linear sigma model \cite{Wi}
(We use the same notation for ${\cal N}=2$ theories 
in two-dimensions as \cite{HoVa}). 
The gauged linear sigma model contains $N+1$ chiral superfields 
$\Phi_i~(i=1,\ldots,N+1)$ and $N-1$ vector superfields
$V_a~(a=1,\ldots,N-1)$ with field strength $\Sigma_a$, which contribute
to the gauge group $U(1)^{N-1}=\prod_{a=1}^{N-1}U(1)_a$. 
The theory is parameterized by Fayet-Iliopoulos (FI) parameter $r_a$, 
theta angle $\theta_a$ and gauge coupling $e_a$. 
We can combine $r_a$ and $\theta_a$ into a complex
parameter $t_a=r_a-i\theta_a$. The Lagrangian is given by
\begin{equation}
  \label{lagrangian}
  L = \int d^4\theta \left( \sum_{i=1}^{N+1}
      \overline{\Phi}_i e^{2\sum_{a=1}^{N-1}Q_{ai}V_a}\Phi_i
      -\sum_{a=1}^{N-1}\frac{1}{2e_a^2}\overline{\Sigma}_a \Sigma_a \right)
      -\frac{1}{2}\left(
      \int d^2\widetilde{\theta} \sum_{a=1}^{N-1}t_a \Sigma_a + {\rm c.c.}
      \right),
\end{equation}
where the $U(1)_a$ charges $Q_{ai}$ carried by $\Phi_i$ are as follows:
\begin{equation}
  \label{charge}
  Q_{ai}=\left\{\begin{array}{rl}
                -2, & i=a+1,\\
                 1, & i=a, a+2,\\
                 0, & {\rm otherwise}.
                \end{array}\right.
\end{equation}
These charges are identified with the charge vectors 
in the toric data for ${\cal M}$.

After integrating out the auxiliary fields, we have the following 
scalar potential energy derived from the Lagrangian (\ref{lagrangian})  
\begin{equation}
  \label{scalarpot}
  V = \sum_{a=1}^{N-1}\left\{ \left( \frac{e_a^2}{2}\sum_{i=1}^{N+1}
      Q_{ai}|\phi_i|^2 - r_a \right)^2 + \sum_{i=1}^{N-1}Q_{ai}^2|\sigma_a|^2
      |\phi_i|^2 \right\},
\end{equation}
where $\phi_i$ and $\sigma_a$ are the scalar components of $\Phi_i$ and
$\Sigma_a$, respectively. 
Let us describe the classical vacuum moduli space which is the space of
zeros of $V$ modulo gauge transformations.
In the case of $r_a \not= 0$, a solution is given by
\begin{equation}
  \label{higgs}
  \sum_{i=1}^{N+1} Q_{ai}|\phi_i|^2 = r_a,~~~~a=1,\ldots,N-1,
\end{equation}
with the gauge identification $\exp(\sum_{a=1}^{N-1}Q_{ai}\gamma_a)\phi_i
\sim \phi_i$ ($\gamma_a \in {\bf R}$) and $\sigma_a=0$.
In this region, if we take the limit $e_a^2 \to \infty$ or
the long distance limit,
the system reduces to the non-linear sigma model whose target
space is the configuration space of classical vacua (\ref{higgs}) 
modulo gauge transformation.
This target space is nothing but ${\cal M}$.
Quantum mechanically, we would have singularities
of the quantum theory which emerge at some specific values of $t_a$.
However, two generic smooth points in the moduli space can be smoothly
connected without passing the singularities.
Thus, ${\cal M}$ can be realized as a vacuum moduli space of
the linear sigma model.
 
The region $r_a>0$ corresponds to the blow-up phase,
where we have the exceptional divisors ${\bf P}^1$ described by 
$\phi_a, \phi_{a+2}$ which cannot vanish simultaneously.
On the other hand, in the region $r_a<0$,
$\phi_{a+1}$ cannot vanish and the exceptional divisors are blown down.
The $N-1$ exceptional divisors $\alpha_a~(a=1,\ldots,N-1)$ 
appearing from the singularity of ${\bf C}^2/{\bf Z}_N$ 
are given by the loci
\begin{equation}
\label{divisor}
  \alpha_a ~:~ \{ \phi_{a+1} = 0 \},~~~~a=1,\ldots,N-1.
\end{equation}
Roughly speaking, the FI parameters $r_a~(a=1,\ldots,N-1)$ control the 
volume of the 2-cycles $\alpha_a$.
On the other hand, $\theta_a$ reduce to $B$ fields in the sigma model.
The reader should note that 
we can rely on the sigma model description only in the blow-up phase.

In the large volume limit ($r_a \to +\infty$), 
the BPS objects in the bulk theory on ${\cal M}$ would be
provided by D0-branes on ${\cal M}$, the bound states of them, 
(anti-) D2-branes wrapping the 2-cycles $\alpha_a$ 
and their bound states with D0-branes. 
In the next section, 
we will take the opposite limit ($r_a \to -\infty$), i.e. orbifold point
where all $N-1$ ${\bf P}^1$ cycles in ${\cal M}$ shrink
and the smooth ALE space ${\cal M}$ becomes 
the singular orbifold ${\bf C}^2/{\bf Z}_N$.
We will see how the BPS states 
in the large volume and orbifold pictures are related to each other. 
This will be carried out by making use of quiver gauge theory \cite{DoMo}.

\section{Quiver gauge theory}
\hspace{5mm}
Consider the gauge theory on the worldvolume of a single D0-brane
probe placed at the orbifold singularity of ${\bf C}^2/{\bf Z}_N$ \cite{DoMo}.
The quiver gauge theory under consideration is a projection of the
ordinary gauge theory on $N$ D0-branes in the covering space 
${\bf C}^2$. 
The resulting field theory is summarized by a quiver diagram. 
Its gauge group is the product $\prod_{i=1}^N U(n_i)$.
The matter content of the quiver gauge theory is $a_{ij}$
bi-fundamental hypermultiplets in the representation $\oplus_{i,j=1}^N
a_{ij}(\bar{n}_i,n_j)$.
Here, this $a_{ij}$ is determined by $C_{ij}=-2\delta_{ij}+a_{ij}$,
where $C_{ij}$ is the Cartan matrix of the affine Lie algebra 
$\widehat{A}_{N-1}$.
Also the theory has $N$ 
Fayet-Iliopoulos (FI) parameters $\zeta_i~(i=1,\dots,N)$. 
Then, the first $N-1$ parameters correspond in the large volume limit
to the blowing up parameters $r_a~(a=1,\dots, N-1)$ 
which appeared in the previous section.
We can remove the $N$-th FI parameter $\zeta_N$ 
by the condition $\sum_{i=1}^N \zeta_i=0$.

In order to specify BPS states, 
we have to solve the F- and D-flatness conditions of the gauge theory.
In ${\cite{FiMa}}$, it was realized that 
a representation theory of a quiver diagram enables us 
to give an efficient way to analyze BPS spectrum.
We will explain how this quiver gauge theory encodes the data about the 
BPS spectrum near the orbifold point.

\subsection{Representation of a quiver diagram}
\hspace{5mm}
Let us begin with a configuration of the quiver gauge theory as 
a representation of the quiver diagram.
The set of non-negative integers $n=(n_1,\ldots,n_N)$, 
which we call dimension vector, specifies the gauge group. 
The representation space with the fixed dimension vector $n$
corresponds to the configuration space of the quiver gauge theory.
We can specify a subrepresentation of a generic representation 
with dimension vector $n$.
The dimension vector of this subrepresentation, 
$n'=(n_1',n_2',\dots,n_N')$, is called a subvector of $n$.

The dimension vectors are in one-to-one correspondence with the lattice
spanned by the simple roots of the affine Lie algebra ${\widehat A}_{N-1}$.
The dimension vector specifies the site
in the positive root lattice $\Gamma_+$ of $\widehat{A}_{N-1}$
\begin{equation}
\label{lattice}
  \Gamma_+ = \left\{\sum_{i=1}^N n_i \alpha_i \right\},
\end{equation}
where $\alpha_i~(i=1,\ldots,N)$ are the simple roots of
$\widehat{A}_{N-1}$. 
The first $N-1$ simple roots $\alpha_i~(i=1,\ldots,N-1)$
correspond to the $N-1$ exceptional divisors in ${\cal M}$. 

Suppose we have a certain BPS state with dimension vector $n$. 
The central charge $Z$ of this state reads \cite{FiMa}
\begin{equation}
Z=\sum_{i=1}^N  \zeta_in_i +i \: \frac{\sum_{i=1}^N n_i}{N}.
\end{equation}
The mass $M$ of BPS states is associated with the central charge $Z$ 
by $M=|Z|/g_s$, where $g_s$ is string coupling constant.
BPS states in ${\cal M}$ would occupy only a subset of the lattice $\Gamma_+$. 
Thus, we have to determine 
which sites in the lattice $\Gamma_+$ (\ref{lattice}) 
are occupied by the BPS states and the degeneracy of each states. 
This is carried out in the following subsection.

\subsection{BPS spectrum}
\hspace{5mm}
Let us introduce the real $N$-dimensional vector 
$\theta=(\theta_1,\theta_2,\dots,\theta_N)$.
The representation with dimension vector $n$ 
is $\theta$ stable if, for any subrepresentation with subvector $n'$
and any $\theta$ vector satisfying $\sum_{i=1}^N n_i\theta_i=0$ 
\cite{DoFiRo, FiMa},
\begin{eqnarray}
\sum_{i=1}^N n_i'\theta_i > 0.
\end{eqnarray}
The importance of this $\theta$ stability lies within
the following mathematical fact: $\theta$ stable representations
are in one-to-one correspondence with the solutions to the
F- and D-flatness conditions.
Thus, $\theta$ stable representations correspond to BPS states.

The vector $\theta$ can be mapped into FI parameters in the quiver
gauge theory.
The relation reads \cite{DoFiRo}
\begin{equation}
\label{FI-map}
\theta_i=\zeta_i-\frac{\sum_{i=1}^Nn_i\zeta_i}{\sum_{i=1}^Nn_i}.
\end{equation}
If we shift the value of FI parameters,
the volume of exceptional divisors changes.
At a certain value, we would encounter
BPS states at threshold.
Let us sketch this situation as follows.
We denote a collection of BPS states by $S^i$ and $S_1^f,\dots,S_m^f$,
whose central charges are $Z^i$ and $Z_1^f,\dots,Z_m^f$.
If the condition
\begin{eqnarray}
\label{marginal}
|Z^i|&=&|Z_1^f|+|Z_2^f|+\dots+|Z_m^f|,\nonumber\\
{\rm arg} Z^i &=& {\rm arg} Z_1^f + {\rm arg} Z_2^f + \dots +{\rm arg} Z_m^f,
\end{eqnarray}
is satisfied in the space of FI parameters, 
the decay of an initial BPS state $S^i$ into final $m$ BPS states 
$S_1^f, S_2^f,\dots,S_m^f$ is allowed.
However, the collection of BPS states 
$S_1^f,\dots,S_m^f$ does not correspond to a
$\theta$ stable representation.
If the condition (\ref{marginal}) is satisfied,
we should see $S^i$ and $S_1^f,\dots,S_m^f$ to be an identical state.

With this kept in mind,
we can proceed by introducing $\theta$ semi-stability. 
The representation with dimension vector $n$ 
is $\theta$ semi-stable if, for any subrepresentation with subvector $n'$
and any real $\theta$ vector satisfying $\sum_{i=1}^N n_i\theta_i=0$,
\begin{eqnarray}
\sum_{i=1}^N n_i'\theta_i \geq 0.
\end{eqnarray}
In general, a $\theta$ semi-stable representation cannot correspond to
any solution to the F- and D-flatness conditions.
However, there is a distinguished $\theta$ semi-stable representation
which corresponds to a solution to the F- and D-flatness conditions.
Actually, this representation is 
a direct sum of $\theta$ stable representations,
which was called a graded representation in \cite{FiMa}.
This representation is none other than what we want.
Furthermore,
the notion of $S$-equivalence enables us to identify 
$S^i$ and $S_1^f,\dots,S_m^f$ if the condition (\ref{marginal}) is satisfied.
We should note that $S$-equivalence excludes all the $\theta$ semi-stable 
representations which do not correspond to solutions
to the F- and D-flatness conditions.

Now, we define the moduli space of BPS states as
the space of $\theta$ semi-stable representations with $S$-equivalence.
We should assure that this moduli space have an non-negative dimension.
Then, the correspondence between the subset of $\Gamma_+$ 
and the BPS spectrum in the large volume limit
is determined as follows \cite{FiMa}.
\bigskip

({\bf I}) The positive root of $A_{N-1}$ given by $\alpha_+$ corresponds to
D2-brane wrapping the 2-cycle $\alpha_+$ in ${\cal M}$.

({\bf II}) The null roots $n \delta~(n\geq1)$ correspond to the bound
states of $n$ D0-branes on ${\cal M}$.

({\bf III}) The roots of the form $\alpha_+ + n\delta~(n \geq 1)$
correspond to the $n$-th KK excitation of D2-brane wrapping the 
2-cycle $\alpha_+$ in ${\cal M}$.

({\bf IV}) The roots of the form $-\alpha_+ + n\delta~(n \geq 1)$
correspond to the $(n-1)$-th KK excitation of anti-D2-brane wrapping
the 2-cycle $\alpha_+$ in ${\cal M}$.
\bigskip

We have introduced above the notations $\alpha_+$ and $\delta$:
\begin{eqnarray}
  \alpha_+ &\in& \{\alpha_{ij} \equiv \alpha_i+\alpha_{i+1}+\cdots+\alpha_j
                 ~;~1 \leq i \leq j \leq N-1\}, \CR
  \delta &=& \sum_{i=1}^N \alpha_i,
\end{eqnarray}
which stand for a positive root of the Lie algebra $A_{N-1}$ and the
null root of $\widehat{A}_{N-1}$, respectively. 
Then, the positive root
$\alpha_+ = \alpha_i+\cdots+\alpha_j$ can be interpreted as the
homological sum of the 2-cycles $\alpha_i,\ldots,\alpha_j$ in (\ref{divisor}). 
We use the same symbol $\alpha_+$ to specify the corresponding
2-cycle in ${\cal M}$.
The dimension vector corresponding to the simple root $\alpha_i$ has
entry one in the $i$-th component with all other entries being zero, 
and one associated with the null vector is $(1,1,\dots,1)$.

\subsection{Marginal stability}
\hspace{5mm}
We give now general remarks on the moduli space of BPS states.
If the condition (\ref{marginal})
is satisfied, the decay of the initial BPS state $S^i$ into
other final $m$ BPS states $S_1^f, S_2^f,\dots,S_m^f$ is allowed.
In terms of the representation of a quiver diagram, 
$\theta$ stable representation becomes $\theta$ semi-stable
representation, which is $S$-equivalent to a direct sum 
of $\theta$ stable subrepresentations. 
In general, the condition (\ref{marginal})
is satisfied in a certain submanifold of the moduli space.
This submanifold is usually called a ``line of marginal stability''.
However, the reader should not confuse the term ``line'' with 
a line in the moduli space.
Actually, the locus of the marginal stability 
is a codimension one submanifold in the moduli space.
On the other hand, due to the celebrated McKay correspondence,
we have a one-to-one correspondence between BPS states
in ${\cal M}$ and root vectors of ${\widehat A}_{N-1}$ 
throughout the moduli space.
This fact means that we cannot miss any BPS states
in wandering the moduli space.
Thus, we always have the lines
of marginal stability, which do not lead to the decay of BPS states. 

Instead of making a general analysis on lines of marginal stability, 
we show an example only for the simplest case, 
${\bf C}^2/{\bf Z}_2$ orbifold.
Consider the $\theta$ stable representation with $n=(1,1)$.
This corresponds to a D0-brane in ${\cal M}$.
We choose $(\theta_1,\theta_2)$ to satisfy the relation 
$\sum_{i=1}^2n_i\theta_i=\theta_1+\theta_2=0$.
At $\theta_1=0$, $\theta$ stable representation $n$ 
becomes $\theta$ semi-stable, 
and $S$-equivalent to the direct sum of two 
subrepresentation with $n'=(1,0)$ and $n'=(0,1)$.
The corresponding marginal stability line $\theta_1=0$ 
is rewritten in terms of the FI parameter as $\zeta_1=0$ using (\ref{FI-map}). 
In other words,
we have the following BPS state at threshold on the marginal stability line
\begin{equation}
\label{decay}
{\rm D0} \longleftrightarrow {\rm D2} + {\overline {\rm D2}}.
\end{equation}
Note that, on both sides of marginal stability line,
D0-brane is $\theta$ stable.
Thus, D0-brane cannot decay anywhere in the moduli space.
In this way, the line of marginal stability can be determined 
by examining the $\theta$ stability, 
not by solving the condition (\ref{marginal}).

One might wonder that the analysis in this section 
cannot be applied to
the large volume region in the moduli space of ${\cal M}$.
Our motivation for using mirror symmetry \cite{HoVa, HoIqVa} 
is to remedy this deficiency.
In the next section, the LG mirror of the sigma model 
on ${\cal M}$ will be studied. 
We will often use there the terms ``D2-brane'' or ``D0-brane'' 
in the sigma model sense to specify BPS states, 
although they are meaningful notations 
only in the large volume limit of the sigma model. 
Furthermore, we remind the reader not to confuse D2-branes
in the sigma model with D2-branes in the LG theory which will appear in
the next section.

\section{Landau-Ginzburg theory}
\hspace{5mm}
The non-linear sigma model in the previous section has a mirror
counterpart which is described in terms of a LG theory 
\cite{HoVa, HoIqVa}. 
The field contents of the mirror LG theory consist of $N+1$ twisted chiral
superfields $Y_i~(i=1,\ldots,N+1)$ and the $N-1$ twisted chiral
superfields $\Sigma_a~(a=1,\ldots,N-1)$ which have already appeared in
the previous section as the super field strengths for $V_a$. The twisted
superpotential reads \cite{HoVa}
\begin{equation}
  \label{superpot1}
  \widetilde{W} = \sum_{a=1}^{N-1}\Sigma_a \left(
              \sum_{i=1}^{N+1}Q_{ai}Y_i-t_a\right)
              +\mu\sum_{i=1}^{N+1}e^{-Y_i},
\end{equation}
where $\mu$ is a cut-off parameter.

Our aim in this section is to study the BPS states and their stability
in the bulk theory. To this end, it suffices to compute the period
integral of the form \cite{CeVa,HoVa}
\begin{equation}
  \label{period}
  \Pi = \int \prod_a d\Sigma_a \prod_i dY_i \exp(-\widetilde{W})
\end{equation}
which, when integrated over appropriate regions in the field
configuration space, yields the BPS masses. The expression of the period
(\ref{period}) can be simplified following the manipulation used in
\cite{HoIqVa}. Integrating out $\Sigma_a$ leads to the constraints among
the superfields $Y_i$,
\begin{equation}
  \label{constraint}
  \sum_{i=1}^{N+1}Q_{ai}Y_i = t_a.
\end{equation}
Let us apply the constraints (\ref{constraint}) to the twisted
superpotential (\ref{superpot1}). If we define two twisted chiral
superfields by
\begin{eqnarray}
  v &=& \mu \exp \left[ -Y_2 - \frac{1}{N}\sum_{i=1}^{N-1}(N-i)t_i \right] \CR
  w &=& \exp \left[ Y_1-Y_2 - \frac{1}{N}\sum_{i=1}^{N-1}(N-i)t_i \right]
\end{eqnarray}
which take their values in ${\bf C}^*$, then all other superfields
$Y_i~(i=3,\ldots,N+1)$ can be expressed by $v$ and $w$ with the help of
the charges (\ref{charge}). The resulting period is
\begin{equation}
  \label{period2}
  \Pi = \int \frac{dvdw}{vw}\exp(-\widetilde{W}),~~~~
  \widetilde{W}=v\left(w^{-1} + \sum_{i=1}^{N-1}c_i w^{i-1} + w^{N-1}\right),
\end{equation}
where the complex parameters $c_i~(i=1,\ldots,N-1)$ are given by
\begin{equation}
  c_i = \exp\left[ \left(1-\frac{i}{N}\right)\sum_{j=1}^{i-1}j t_j
                  +\frac{i}{N}\sum_{j=i}^{N-1}(N-j)t_j \right].
\end{equation}
In (\ref{period2}) and hereafter, we will omit unimportant numerical 
factors in front of the period. The variable $v$ cannot be simply
integrated out since it takes values in ${\bf C}^*$. In other words, the
integral measure in (\ref{period2}) contains the factor $dv/v$. To
appropriately eliminate $v$, we introduce two additional twisted chiral
superfields $x$ and $y$ which take values in ${\bf C}$. Then, it is easy
to verify that the period (\ref{period2}) is equivalent to
\begin{equation}
  \label{period3}
  \Pi = \int \frac{dvdwdxdy}{w} \exp(-\widetilde{W}),~~~~
  \widetilde{W}= v\left( w^{-1} + \sum_{i=1}^{N-1}c_i w^{i-1} + w^{N-1}
                 +xy \right),
\end{equation} 
where $v$ can be viewed as a ${\bf C}$ variable. In fact,
integrating $x$ and $y$ out reproduces (\ref{period2}) and 
especially the measure $dv/v$, 
which ensures that $v$ was originally a ${\bf C}^*$ variable. 

The variable $v$ can now be safely integrated out, leading to the
following expression of the period,
\begin{equation}
  \label{period4}
  \Pi = \int \frac{dwdxdy}{w}
  \delta\left(w^{-1} + \sum_{i=1}^{N-1}c_i w^{i-1} + w^{N-1} + xy \right).
\end{equation}
This result implies that we can analyze the BPS spectrum by identifying
the LG theory with the sigma model on the mirror CY $\widehat{{\cal M}}$
given by
\begin{equation}
  \label{cy2}
  \widehat{{\cal M}} ~:~
    \left\{\begin{array}{l}
             f(w,z) \equiv 1 + zw + \sum_{i=2}^{N-1}c_i w^i + w^N = 0, \\
             xy = z-c_1,
           \end{array}\right.
\end{equation}
where we have introduced an extra ${\bf C}$ variable $z$. The final
expression of the period is obtained by integrating out $y$ in
(\ref{period4}) as
\begin{equation}
  \label{period5}
  \Pi = \int \Omega,~~~~
  \Omega = \frac{dwdx}{wx}.
\end{equation}
The 2-form $\Omega$ defined above is none other than the holomorphic
2-form on $\widehat{{\cal M}}$.

The geometry of $\widehat{{\cal M}}$ takes the form of the product of two
fibrations over the base $B \simeq {\bf C}$ parameterized by $z$. One of
the two fibers is the set of $N$ points $F_1=\{w_1(z),\ldots,w_N(z)\}$
in the $w$ plane, which solve the equation $f(w,z)=0$. Let us define the
discriminant of the polynomial $f(w,z)$ by $\Delta(z) \equiv
\prod_{i-1}^N(z-z_i)$. Then, some two points in $F_1$ coalesce when $z$
meets one of the $N$ points $P=\{z_1,\ldots,z_N\}$ on $B$. The other fiber
is the algebraic torus $F_2 \simeq {\bf C}^*$ given by $xy=z-c_1$, which
degenerates at $\{z=c_1\}$.

\subsection{Special Lagrangian submanifolds}
\begin{figure}
    \centerline{\psfig{figure=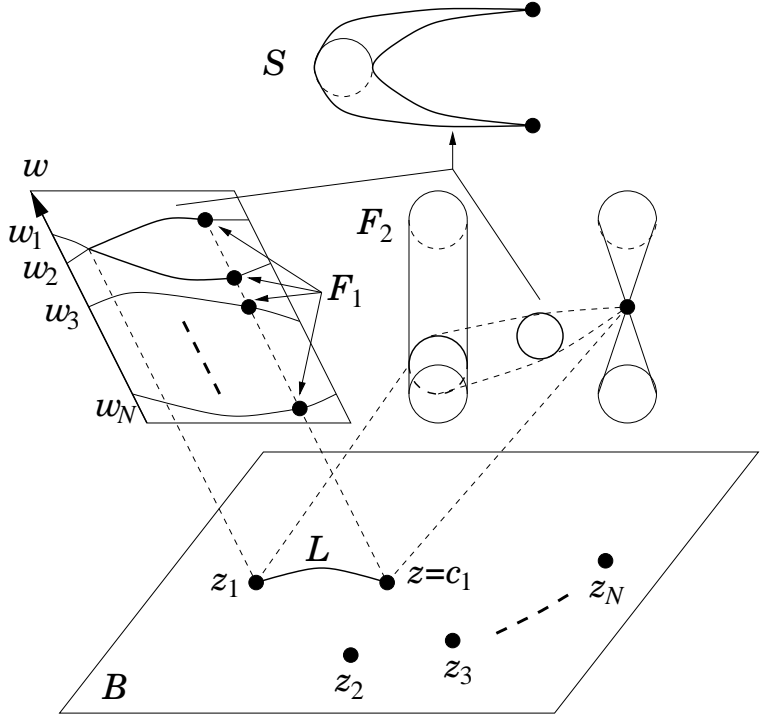,width=10cm}}
    \caption{Non-compact mirror Calabi-Yau $\widehat{{\cal M}}$ and a
 2-sphere $S$ embedded in it.}
  \label{cyfig}
\end{figure}

\hspace{5mm}
We have to look for all the special Lagrangian submanifolds in
$\widehat{{\cal M}}$, which contribute to the BPS states in the bulk
theory. Before identifying the special Lagrangian
submanifolds, let us consider for an exercise how middle dimensional
compact submanifolds, which are not necessarily special Lagrangian, are
embedded in $\widehat{{\cal M}}$. A simple example would be 2-sphere, as
we will explain now. First consider a line interval $L$ on $B$
connecting $\{z=c_1\}$ and one of the $N$ points in $P$, say
$\{z=z_1\}$. Then, a 2-sphere $S$ can be constructed as a submanifold
fibered over $L$. The real one dimensional fiber $f$ of $S$ lies in the
fiber of $\widehat{{\cal M}}$, $F_1 \times F_2$. The projection of $f$
on $F_1$ consists of the two points in $F_1$, which meet each other at
the endpoint $\{z=z_1\}$ of $L$. On the other hand, the projection of
$f$ on $F_2 \simeq {\bf C}^*$, which can be interpreted as an infinitely 
elongated cylinder, is the compact circle in the cylinder which shrinks
at $\{z=c_1\}$. It is evident that $S$ obtained this way forms a
2-sphere in $\widehat{{\cal M}}$. It is shown in Figure \ref{cyfig} how
the 2-sphere $S$ is embedded in $\widehat{{\cal M}}$.

Let us turn to the problem of finding the special Lagrangian
submanifolds in $\widehat{{\cal M}}$. For a submanifold $C$ to be
special Lagrangian, the following conditions must be satisfied
\cite{BeBeSt,Joyce,KaMc}:
\begin{eqnarray}
  \omega|_C &=& 0, \label{slagk} \\
  {\rm Im}(e^{-i\theta}\Omega)|_C &=& 0. \label{slagh}
\end{eqnarray}
Here, $\omega$ is the K\"{a}hler form on $\widehat{{\cal M}}$,
\begin{equation}
  \label{kahler}
  \omega = \frac{i}{2}(dwd\bar{w} + dxd\bar{x} + dyd\bar{y}),
\end{equation}
where $f(w,xy+c_1)=0$ is implicitly assumed to eliminate $y$. The phase
$\theta \in [0,2\pi)$
\footnote{
In \cite{DoFiRo}, it has been argued that this phase should be extended
to any real number ${\bf R}$. This leads us to introduce the notion
``graded Lagrangian submanifold'' as in \cite{Seidel}.
}
should be constant over all points on $C$. We
often call $C$ a special Lagrangian submanifold with phase $\theta$.

The form of $\Omega=d(\ln w)d(\ln x)$ 
and the condition (\ref{slagh}) suggest that
a compact special Lagrangian submanifold projected on both the $(\ln
w)$- and $(\ln x)$- plane must be a straight line. To specify how the
straight line is lying on the $(\ln w)$-plane, it is convenient to
introduce an infinite number of points,
\begin{equation}
  \label{points}
  u_i(n) = \ln w_i(c_1) +2\pi in,~~~~i=1,\ldots,N,~~n \in {\bf Z},
\end{equation}
where  $0 \leq {\rm Im}\ln w_i(c_1) < 2\pi$ is assumed.
There are two possibilities for the form of the straight line on the
$(\ln w)$-plane. If the straight line is the line interval connecting
$u_i(n)$ and $u_j(m)$ with $i \neq j$, the associated special Lagrangian
submanifold has the topology of 2-sphere. In fact, we can parameterize
the submanifold by parameters $s,t \in [0,1]$ as
\begin{eqnarray}
  \label{sphere}
  \ln w &=& u_i(n)(1-t)+u_{j}(m)t, \CR
  \ln x &=& \frac{1}{2}\ln \left| \frac{f(w,c_1)}{w} \right| + 2\pi i s,
\end{eqnarray}
where we have obtained the second line by examining the condition
$(\ref{slagk})$. Keep in mind that $w$ in the second line must be
replaced by a function of $t$ using the first line. The
variables $t$ and $s$ parameterize a line interval and a circle which is
fibered over the interval and shrinks at the endpoints of it,
respectively. The special Lagrangian submanifolds of this type belong to
the same homology class as that of the 2-sphere $S$ mentioned above.

On the other hand, we can imagine that the straight line on the $(\ln
w)$-plane starts from an arbitrary real number $a \in {\bf R}$ and ends
on $a+2\pi i$. Similarly to the above case, the form of the straight
line on the $(\ln x)$-plane can be determined from (\ref{slagk}). The
result is
\begin{eqnarray}
  \label{torus}
  \ln w &=& a(1-t)+(a+2\pi i)t, \CR
  \ln x &=& \frac{1}{2}\ln \left[ b +
            \sqrt{b^2+\left|\frac{f(w,c_1)}{w}\right|^2} \right]
            + 2\pi i s,
\end{eqnarray}
where $b$ is an arbitrary real constant. Again we have to use the
first line to eliminate $w$ in the second line. In this case, the
associated special Lagrangian submanifold is a torus. Notice that this
special Lagrangian torus has two real dimensional moduli $a$ and $b$,
while special Lagrangian 2-spheres have no moduli. This reflects the
fact that, in general, a compact special Lagrangian submanifold $C$ in a
CY manifold has a deformation moduli space of real dimension $b^1(C)$,
the first Betti number \cite{Joyce}.

\begin{figure}
    \centerline{\psfig{figure=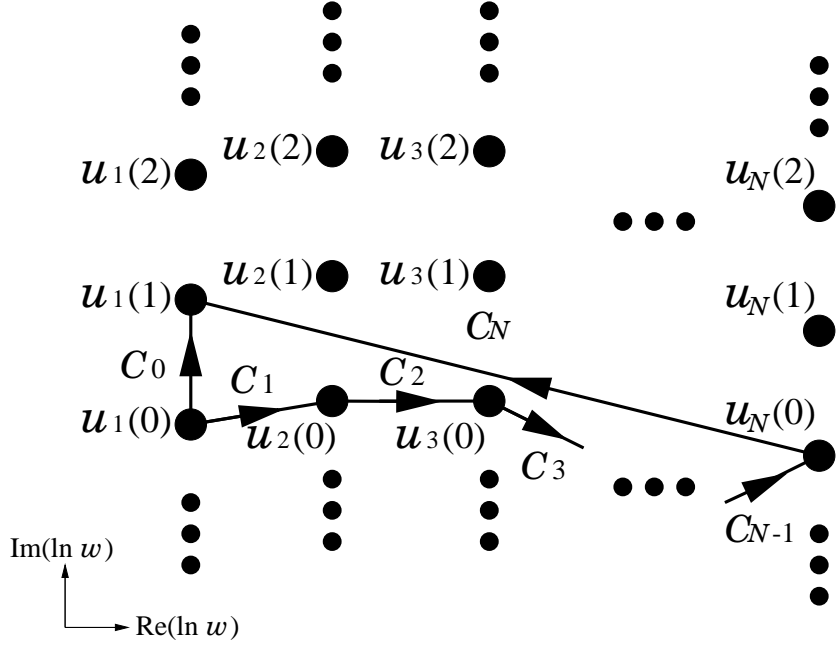,width=11cm}}
    \caption{Special Lagrangian submanifolds in $\widehat{{\cal M}}$
 projected on the $(\ln w)$-plane.}
  \label{slagN}
\end{figure}

We can choose $N$ special Lagrangian 2-spheres $C_i~(i=1,\ldots,N)$, so
that their homology classes $[C_i]$ form a basis of the group of the
compact 2-cycles in $\widehat{{\cal M}}$, $H_2(\widehat{{\cal M}},{\bf
Z})$. The straight line intervals, which are the projection of $C_i$ on
the $(\ln w)$-plane, are given by
\begin{eqnarray}
  \label{interval}
    C_i &:& \ln w = u_i(0)(1-t)+u_{i+1}(0)t,~~~~i=1,\ldots,N-1, \CR
  C_{N} &:& \ln w = u_N(0)(1-t)+u_1(1)t.
\end{eqnarray}
We show in Figure \ref{slagN} the intervals associated with $C_i$.
The adjacent two submanifolds $C_i$ and $C_{i+1}$ intersect only at a
single point $\{w=w_{i+1}(c_1),z=c_1,x=y=0\}$ for $i=1,\ldots,N-1$, and
the unique intersection point between $C_1$ and $C_N$
is given by $\{w=w_{1}(c_1),z=c_1,x=y=0\}$.
Then we can give the homology group $H_2(\widehat{{\cal
M}},{\bf Z})$ a structure of lattice by the intersection numbers,
\begin{equation}
  \label{intersection}
  (C_i \cdot C_j) = \left(
                   \begin{array}{rrrrrr}
                        -2, &     1, &    0,  & \cdots &     0, &      1 \\
                         1, &    -2, &    1,  & \ddots & \ddots &      0 \\
                         0, &     1, & \ddots & \ddots & \ddots & \vdots \\
                     \vdots & \ddots & \ddots & \ddots &     1, &      0 \\
                         0, & \ddots & \ddots &     1, &    -2, &      1 \\
                         1, &     0, & \cdots &     0, &     1, &     -2
                   \end{array}\right).
\end{equation}
Here we have a special 2-cycle given by the homological sum of all the
$N$ 2-cycles, 
\begin{equation}
  \label{c0}
  C_0=\sum_{i=1}^N C_i. 
\end{equation}
As the intersection matrix (\ref{intersection}) indicates, $(C_0 \cdot
C_i)=0$ for $i=1,\ldots,N$. This can be geometrically understood by
identifying $C_0$ with a submanifold which admits a fibration over a
sufficiently large circle on $B$ surrounding all $N$ points in $P$. The
fiber of $C_0$ is a compact circle wrapped on the cylinder $F_2 \simeq
{\bf C}^*$. Therefore, the submanifold $C_0$ has the topology of torus,
whereas $C_i~(i=1,\ldots,N)$ has the shape of 2-sphere. The special
Lagrangian torus given by (\ref{torus}) is in the homology class $C_0$.

\subsection{BPS spectrum}
\hspace{5mm}
In the original sigma model, the BPS spectrum is provided by D0-branes,
(anti-) D2-branes wrapped on holomorphic rational curves in ${\cal M}$
and their KK excitations. These BPS objects are mapped, in the LG
mirror, to the D2-branes wrapped on special Lagrangian submanifolds in
$\widehat{\cal M}$. The special Lagrangian submanifolds can be
represented by linear combinations of $N$ submanifolds $C_i$ given
above. In fact the root lattice of the affine Lie algebra
$\widehat{A}_{N-1}$ is identified with the homology lattice generated by
$C_i$. The correspondence between the roots and the submanifolds $C_i$ is
simply given by
\begin{eqnarray}
  \alpha_i &\leftrightarrow& C_i,~~~~i=1,\ldots, N, \CR
  \alpha_+ &\leftrightarrow& C_+ \in \{C_{ij}=C_i+C_{i+1}+\cdots+C_j~;~
                             1 \leq i \leq j \leq N-1\}, \CR
    \delta &\leftrightarrow& C_0.
\end{eqnarray}
The BPS objects in the sigma model on ${\cal M}$, which are classified
by the positive roots of $\widehat{A}_{N-1}$, are therefore in one-to-one
correspondence with D2-branes wrapped on special Lagrangian submanifolds
in $\widehat{{\cal M}}$ as follows.
\bigskip

({\bf I}) D2-brane wrapped on $\alpha_{ij}$ in ${\cal M}$ is
mapped to D2-brane wrapped on $C_{ij}$ in $\widehat{{\cal M}}$.

({\bf II}) The bound state of $n~(\geq 1)$ D0-branes on ${\cal M}$ is
mapped to D2-brane wrapped on $nC_0$ in $\widehat{{\cal M}}$.

({\bf III}) The $n~(\geq 1)$-th KK excitation of D2-brane wrapped on
$\alpha_{ij}$ in ${\cal M}$ corresponds to D2-brane wrapped on $C_{ij} +
n C_0$ in $\widehat{{\cal M}}$.

({\bf IV}) The $n~(\geq0)$-th KK excitation of anti-D2-brane wrapped
on $\alpha_{ij}$ in ${\cal M}$ is identified with D2-brane
wrapped on $-C_{ij} + (n+1) C_0$ in $\widehat{{\cal M}}$.
\bigskip

The mass of the BPS state, which corresponds to the D2-brane wrapped on a
special Lagrangian submanifold $C$ of $\widehat{{\cal M}}$, is given by
\begin{equation}
  \label{BPSmass}
  M=\frac{1}{4\pi^2g_s}\left|\int_C \Omega\right|.
\end{equation}
In this formula, the normalization factor is included to reproduce the
results in the large volume limit which will be presented below. It is
rather difficult for an arbitrary $N$ to obtain the BPS mass formula in
a closed form by computing (\ref{BPSmass}). We will perform a detailed
analysis only for the simplest cases $N=2,3$ in the following
subsections. Before doing it, however, it would be worthwhile to confirm
that our mirror model reproduces some known results for general $N$ in
the two extremal situations: the large volume and orbifold
limits. Specifically, we will focus on $C_i~(i=1,\ldots,N-1)$ and
evaluate the complexified K\"{a}hler class of their mirror
\begin{equation}
  \label{kahler-class}
  B_{\alpha_i}+iJ_{\alpha_i} = \frac{1}{4\pi^2} \int_{C_i} \Omega,
\end{equation}
in these limits.
Here, $J_{\alpha_i}$ and $B_{\alpha_i}$ denote respectively the 
K\"ahler form and $B$ field on ${\cal M}$ integrated over the 2-cycle
$\alpha_i$.
\bigskip

\underline{{\large Large volume limit}}
\bigskip

The large volume limit is to take ${\rm Re}~t_a = r_a \rightarrow
+\infty$. Let us study how the K\"{a}hler class (\ref{kahler-class})
behaves in this limit. For ${\rm Re}~t_a \gg 1$, the $N$ solutions
$\{w_1(c_1),\ldots,w_N(c_1)\}$ to the equation $f(w,c_1)=0$ are
approximately given by
\begin{eqnarray}
  \label{solution}
  w_1(c_1) &\sim& 1/c_1, \CR
  w_i(c_1) &\sim& c_{i-1}/c_{i},~~~~i=2,\ldots,N-1, \CR
  w_N(c_1) &\sim& c_{N-1}.
\end{eqnarray}
This estimation was derived by assuming $|w_1(c_1)| \ll |w_2(c_1)| \ll
\cdots \ll |w_N(c_1)|$. Then, the K\"{a}hler class of the holomorphic 2-cycles
in ${\cal M}$ mirror to $C_i$ can be read off by combining
(\ref{period5}), (\ref{points}), (\ref{interval}), (\ref{BPSmass}) and
(\ref{solution}) as
\begin{eqnarray}
  B_{\alpha_i}+iJ_{\alpha_i} &=& \frac{i}{2\pi}
                                 \ln\frac{w_{i+1}(c_1)}{w_i(c_1)}, \CR
                          &\sim& \frac{i}{2\pi}t_i.
\end{eqnarray}
This reproduces the classical picture that $it_i=\theta_i+ir_i$ can be
interpreted as the complexified K\"{a}hler class of the exceptional
divisor $\alpha_i$.
\bigskip

\underline{{\large Orbifold limit}}
\bigskip

The orbifold limit is realized by taking ${\rm Re}~t_a = r_a
\rightarrow -\infty$. It is interesting to take this limit and see the
fate of the K\"{a}hler class. The $N$ solutions
$\{w_1(c_1),\ldots,w_N(c_1)\}$ in the limit ${\rm Re}~t_a \rightarrow
-\infty$ are given by
\begin{equation}
  \label{solution2}
  w_i(c_1) = \exp\left[\frac{(2i-1)\pi i}{N} \right],~~~~i=1,\ldots,N.
\end{equation}
As in the large volume limit, let us confine ourselves to the D2-brane
wrapping $C_i$. The resultant K\"{a}hler class is
\begin{eqnarray}
  \label{fractional}
  B_{\alpha_i}+iJ_{\alpha_i} &=& \frac{i}{2\pi}
                                 \ln\frac{w_{i+1}(c_1)}{w_i(c_1)}, \CR
                             &=& -\frac{1}{N}.
\end{eqnarray}
We have again succeeded to rederive well known facts at the orbifold
point. Namely, it was verified in (\ref{fractional}) that the $B$ field
takes the value $-1/N$ \cite{AsGrMo,Douglas} and consequently the BPS
mass is smaller than the ordinary unit $1/g_s$ for D0-brane by the
factor $1/N$. Therefore, we can conclude that the D2-brane wrapping
$C_i~(i=1,\ldots,N-1)$ becomes a fractional brane \cite{DiDoGo,DiGo} in the
orbifold limit.

\subsection{${\bf C}^2/{\bf Z}_2$}
\begin{figure}
    \centerline{\psfig{figure=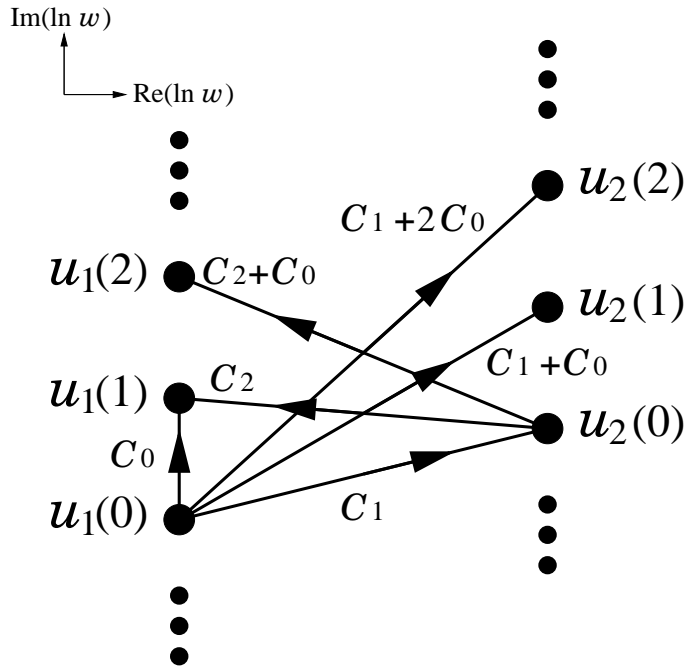,width=9cm}}
    \caption{Special Lagrangian submanifolds projected on the $(\ln w)$-plane.}
  \label{slag1}
\end{figure}

\hspace{5mm}
Let us begin with the simplest example, ${\bf C}^2/{\bf Z}_2$. For this
case, the defining equation (\ref{cy2}) of $\widehat{{\cal M}}$ is given by
\begin{equation}
  \label{c2z2}
  f(w,z)=w^2+zw+1=0,~~~~xy=z-e^{t_1/2}.
\end{equation}
The discriminant $\Delta(z)$ of the first equation in (\ref{c2z2}) is
\begin{equation}
  \Delta(z)=(z-z_1)(z-z_2),~~~~z_1=-2,~z_2=2.
\end{equation}
The special Lagrangian submanifolds in $\widehat{{\cal M}}$ can be found 
by the general procedure explained in subsection 4.1. 
The two solutions of $f(w,e^{t_1/2})=0$ are given by
\begin{equation}
  w_1(e^{t_1/2})=\frac{1}{2}\left(e^{t/2}+\sqrt{e^t-4}\right),~~~~
  w_2(e^{t_1/2})=\frac{1}{2}\left(e^{t/2}-\sqrt{e^t-4}\right).
\end{equation}
These two points in the $w$-plane are mapped to an infinite number of
points aligned on the imaginary axis of the $(\ln w)$-plane,
\begin{equation}
  u_{1,2}(n) = \ln w_{1,2}(e^{t_1/2}) +2\pi in,~~~~n \in {\bf Z},
\end{equation}
where we can assume $0 \leq {\rm Im} [\ln w_1(e^{t_1/2})] \leq {\rm Im}
[\ln w_2(e^{t_1/2})] < 2\pi$ without loss of generality. Then, we
associate the special Lagrangian submanifold $C_1~(C_2=-C_1+C_0)$, which 
corresponds to a (anti-) D2-brane on ${\cal M}$, with the straight line
segments on the $(\ln w)$-plane,
\begin{eqnarray}
  \label{segment-d2}
  C_1 &:& \ln w = u_1(0)(1-t)+u_2(0)t, \CR
  C_2 &:& \ln w = u_2(0)(1-t)+u_1(1)t.
\end{eqnarray}
Other special Lagrangian submanifolds, which give rise to the
$n~(\geq1)$-th KK modes, are given by
\begin{eqnarray}
  \label{segment-kk}
      C_1+n C_0 &:& \ln w = u_1(0)(1-t)+u_2(n)t, \CR
  -C_1+(n+1)C_0 &:& \ln w = u_2(0)(1-t)+u_1(n+1)t.
\end{eqnarray}
All other line segments are identical with the ones appearing in
(\ref{segment-d2}) and (\ref{segment-kk}). Finally, the special
Lagrangian submanifold $nC_0$, which is identified with the bound
state of $n$ D0-branes on ${\cal M}$, is parameterized by
\begin{equation}
  \label{d0}
  nC_0 ~:~ \ln w = a + 2\pi i n t,
\end{equation}
where $a \in {\bf R}$ describes the moduli of deforming
$C_0$ in $\widehat{{\cal M}}$ along the $(\ln w)$-plane. Recall that
there exists one more modulus $b$ as in (\ref{torus}). When combined
with the real two dimensional moduli space of the Wilson line on a
D2-brane wrapping the 2-torus $C_0$, the two parameters $a$ and $b$
account for the real four dimensional moduli space of a D0-brane moving
on ${\cal M}$. In Figure \ref{slag1}, we depict these special Lagrangian
submanifolds projected on the $(\ln w)$-plane.

The mass of a (anti-) D2-brane wrapped on the holomorphic curve
$\alpha_1$ in ${\cal M}$ can then be determined by (\ref{segment-d2}) as
\begin{eqnarray}
  {\rm D2}
    &:& M = \frac{1}{2\pi g_s}\left|\ln\left(\frac{e^{t_1/2}-\sqrt{e^{t_1}-4}}
            {e^{t_1/2}+\sqrt{e^{t_1}-4}}\right)\right|, \CR
  {\rm \overline{D2}}
    &:& M = \frac{1}{2\pi g_s}\left|\ln\left(\frac{e^{t_1/2}-\sqrt{e^{t_1}-4}}
            {e^{t_1/2}+\sqrt{e^{t_1}-4}}\right) - 2\pi i\right|.
\end{eqnarray}
This is in agreement, as expected, with the result of \cite{AsGrMo} in
which the GKZ system of differential equations has been used to compute
the period (\ref{period5}). Similarly, we find from (\ref{segment-kk})
that the bound state of $n$ D0-branes with these states has the mass
\begin{eqnarray}
  {\rm D2}/n{\rm D0}
    &:& M = \frac{1}{2\pi g_s}\left|\ln\left(\frac{e^{t_1/2}-\sqrt{e^{t_1}-4}}
            {e^{t_1/2}+\sqrt{e^{t_1}-4}}\right) + 2\pi ni \right|, \CR
  {\rm \overline{D2}}/n{\rm D0}
    &:& M = \frac{1}{2\pi g_s}\left|\ln\left(\frac{e^{t_1/2}-\sqrt{e^{t_1}-4}}
            {e^{t_1/2}+\sqrt{e^{t_1}-4}}\right) - 2\pi (n+1)i \right|.
\end{eqnarray}
The mass of the bound state of $n$ D0-branes can be read off
from (\ref{d0}),
\begin{equation}
  n{\rm D0} ~:~ M = \frac{n}{g_s},
\end{equation}
which is a familiar result.

\begin{figure}
    \centerline{\psfig{figure=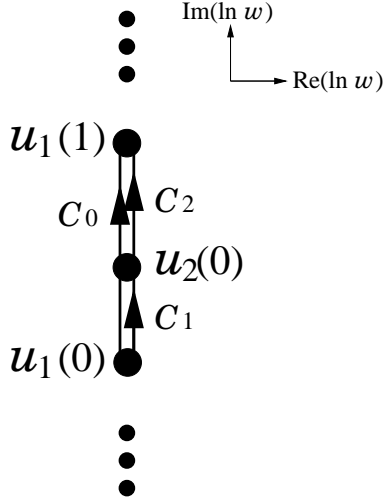,width=5cm}}
    \caption{Special Lagrangian submanifolds on the marginal stability line.}
  \label{decay1}
\end{figure}

We can use the data of special Lagrangian submanifolds to examine the
stability of the BPS states. Let us consider the decay of a BPS state
$S^i$ into other $n$ BPS states $S^f_1,S^f_2,\ldots,S^f_n$. We denote
the special Lagrangian submanifolds in $\widehat{{\cal M}}$ which
correspond to these states by $C^i$ and $C^f_1,C^f_2,\ldots,C^f_n$,
respectively. Furthermore, let $\theta^i$ and
$\theta^f_1,\ldots,\theta^f_n$ be the special Lagrangian phases of
them. For this decay to be possible, the following geometrical condition
is required to hold:
\begin{eqnarray}
  \label{condition}
  C^i &=&  C^f_1+C^f_2+\cdots+C^f_n, \CR
  \theta^i &=& \theta_1^f = \cdots = \theta_n^f.
\end{eqnarray}
The first equation guarantees RR charge conservation. The second
condition states that this decay process is energetically allowed, since
it implies that the mass of the initial state $S^i$ is equal to the sum
of the masses of the final states $S^f_1,S^f_2,\ldots,S^f_n$. In the
present case, there exist submanifolds such that the condition
(\ref{condition}) is satisfied if and only if $\ln w_1(e^{t_1/2})$ and $\ln
w_2(e^{t_1/2})$ have the same real part, as shown in Figure
\ref{decay1}. This condition is rephrased in terms of the parameter
$t_1$ as
\begin{equation}
  \label{line}
  z_1 \leq {\rm Re} ~e^{t_1/2} \leq z_2,~~~~{\rm Im} ~e^{t_1/2} = 0.
\end{equation}

On this line, the decay of the initial state $S^i={\rm D0}$ into the two
final states $S^f_1={\rm D2}$ and $S^f_2=\overline{{\rm D2}}$ is
allowed.
\footnote{
The notation $S^i={\rm D0}$ might be misleading, since the BPS states 
are in one-to-one correspondence with the elements of the $L^2$
cohomology group $H_{L^2}^*(M)$ of a BPS D-brane moduli space $M$
\cite{HaMo,FiMa}, not with the D-branes themselves. However, the moduli
space of D0-brane is the ALE space ${\cal M}$ itself, and
$H_{L^2}^*({\cal M})$ has a unique element given by (the Poincar\'{e}
dual of) the 2-cycle $\alpha_1$ in the case of ${\bf C}^2/{\bf Z}_2$. It
is thus justified in this subsection to adopt that notation. In an
analogous way, we can use D2, D2/D0 and etc. as the names of BPS states.
}
More generally, the bound state of a D2 state and $n$ D0 states
can decay into $n+1$ D2 states and $n$ $\overline{{\rm D2}}$ states. In
summary, the following transition processes can occur on the line of
marginal stability given by (\ref{line}),
\begin{eqnarray}
             {\rm D0} &\leftrightarrow& {\rm D2} + {\rm \overline{D2}}, \CR
  {\rm D2}/{\rm D0} 
     &\leftrightarrow& 2 {\rm D2} + {\rm \overline{D2}}, \CR
     &\vdots& \CR
  {\rm D2}/n{\rm D0} 
     &\leftrightarrow& (n+1) {\rm D2} + n {\rm \overline{D2}},
                        \label{threshold} \\
     &\vdots&. \nonumber
\end{eqnarray}
As stressed in the previous section, on both sides of the line, the
states on the l.h.s. of (\ref{threshold}) are stable. In this regard,
our model, which possesses 16 supercharges, is clearly distinguished
from theories with 8 unbroken supersymmetries where the BPS spectrum can 
jump.
\subsection{${\bf C}^2/{\bf Z}_3$}
\hspace{5mm}
Let us proceed to the ${\bf C}^2/{\bf Z}_3$ orbifold.
The equation (\ref{cy2}) of ${\widehat {\cal M}}$ reads
\begin{equation}
f(w,z)=w^3+e^{\frac{1}{3}t_1+\frac{2}{3}t_2}w^2+zw+1=0,
\qquad xy=z-e^{\frac{2}{3}t_1+\frac{1}{3}t_2}.
\end{equation}

\begin{figure}
    \centerline{\psfig{figure=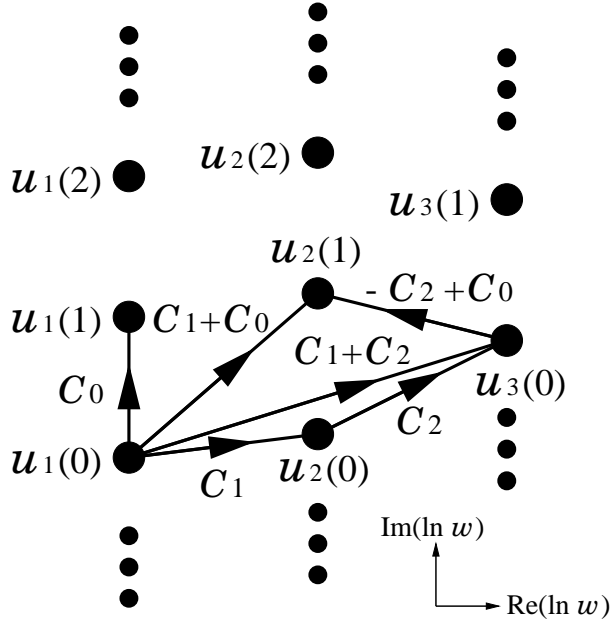,width=8cm}}
    \caption{Special Lagrangian submanifolds projected on the ($\ln w$)-plane}
  \label{slag2}
\end{figure}

Special Lagrangian submanifolds in ${\widehat {\cal M}}$ are
constructed following the general procedure in subsection 4.1.
We write three solutions of
$f(w,e^{\frac{2}{3}t_1+\frac{1}{3}t_2})=0$
by $w_i(e^{\frac{2}{3}t_1+\frac{1}{3}t_2})$ with $i=1,2,3$.
We map these three points in the $w$-plane into an infinite number of
points aligned on the imaginary axis of the $(\ln w)$-plane,
\begin{equation}
u_i(n)=\ln w_i(e^{\frac{2}{3}t_1+\frac{1}{3}t_2})+2\pi i n,\qquad 
n \in {\bf Z},
\end{equation}
where we assume 
$0 \leq {\rm Im}[\ln w_1(e^{\frac{2}{3}t_1+\frac{1}{3}t_2})]
\leq {\rm Im}[\ln w_2(e^{\frac{2}{3}t_1+\frac{1}{3}t_2})]
\leq {\rm Im}[\ln w_3(e^{\frac{2}{3}t_1+\frac{1}{3}t_2})] < 2\pi$.
Then, the special Lagrangian 2-spheres $C_1, C_2$, which correspond to
D2-branes wrapped around holomorphic rational curves 
$\alpha_1, \alpha_2$ in ${\cal M}$, are associated with the 
straight line segments on the $(\ln w)$-plane,
\begin{eqnarray}
  C_1 &:& \ln w = u_1(0)(1-t)+u_2(0)t, \CR
  C_2 &:& \ln w = u_2(0)(1-t)+u_3(0)t.
\end{eqnarray}
The special Lagrangian 2-torus $nC_0$ is parameterized exactly in the
same way as (\ref{d0}).
We depict these special Lagrangian submanifolds in Figure \ref{slag2}.
Other special Lagrangian 2-spheres can be identified easily 
in Figure \ref{slag2}.

Let us study the stability of BPS states in ${\cal M}$ 
using the data of special Lagrangian 2-cycles in ${\widehat {\cal M}}$.
In particular, let us concentrate 
on the bound state of a D2-brane wrapping $\alpha_1$ and a D0-brane.
This state, which we denote by D2$_1$/D0,  
is associated with special Lagrangian 2-cycle
$C_1+C_0$ in ${\widehat {\cal M}}$.
Then, let us consider the following two transition processes
\begin{eqnarray}
\label{decay2}
(a) ~~:~~
{\rm D2_1/D0} &\leftrightarrow& 2{\rm D2_1} + {\rm \overline{D2}_1},\CR
(b) ~~:~~
{\rm D2_1/D0} &\leftrightarrow& {\rm D2_{1+2}} + {\rm \overline{D2}_2}.
\end{eqnarray}
Here, D2$_{1+2}$ denotes a D2-brane wrapping the holomorphic rational
curve $\alpha_1+\alpha_2$.
The decay $(a)$ in $(\ref{decay2})$ is allowed if
$\ln w_1(e^{\frac{2}{3}t_1+\frac{1}{3}t_2})$ and
$\ln w_2(e^{\frac{2}{3}t_1+\frac{1}{3}t_2})$ have the same real part,
\begin{equation}
\label{line1}
{\rm Re}~ \ln \frac{w_2}{w_1}(e^{\frac{2}{3}t_1+\frac{1}{3}t_2})=0.
\end{equation}
On the other hand, the decay $(b)$ is allowed if 
$u_3(0)$ sits on the straight line stretched between $u_1(0)$ and
$u_2(1)$ in the $(\ln w)$-plane,
\begin{equation}
\label{line2}
\ln \frac{w_2}{w_1}(e^{\frac{2}{3}t_1+\frac{1}{3}t_2})
+2\pi i = r \left[
\ln \frac{w_3}{w_2}(e^{\frac{2}{3}t_1+\frac{1}{3}t_2})
-2\pi i \right],~~~~~~~~~~r\in {\bf R}.
\end{equation}
These conditions give marginal stability lines as
real three-dimensional submanifolds in the real
four-dimensional moduli space spanned by two complex FI parameters $t_1,t_2$.

Finally, we wish to show an instructive
analysis on the marginal stability lines
near the orbifold limit $\zeta_1, \zeta_2 \to -\infty$.
\begin{figure}
    \centerline{\psfig{figure=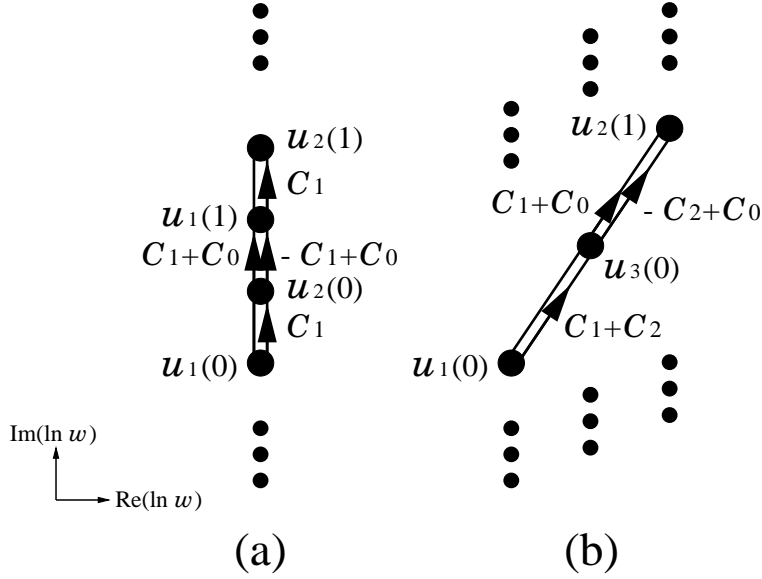,width=10cm}}
    \caption{Special Lagrangian submanifolds on the marginal stability
 line near the orbifold point: (a) $(C_1+C_0) \leftrightarrow 2C_1+(-C_1+C_0)$,
 (b) $(C_1+C_0) \leftrightarrow (C_1+C_2) + (-C_2+C_0)$}
  \label{decays}
\end{figure}
As we have shown in subsection 4.2,
we have $B_{\alpha_1}=B_{\alpha_2}=-\frac{1}{3},~J_{\alpha_1}=J_{\alpha_2}=0$ 
in this limit.
Let us consider the moduli space spanned by $\zeta_i$ 
with the value of $B_{\alpha_i}$ fixed.
Then, we have to notice the following parameterization
\begin{eqnarray}
\label{r-param}
\frac{2\pi}{3}={\rm Im}~\ln w_2(c_1)
-{\rm Im}~\ln w_1(c_1),&& 
\frac{2\pi}{3}={\rm Im}~\ln w_3(c_1)
-{\rm Im}~\ln w_2(c_1),\CR
2\pi \zeta_1={\rm Re}~\ln w_2(c_1)
-{\rm Re}~\ln w_1(c_1),&&
2\pi \zeta_2={\rm Re}~\ln w_3(c_1)
-{\rm Re}~\ln w_2(c_1).
\end{eqnarray}
Here, we have abbreviated $e^{\frac{2}{3}t_1+\frac{1}{3}t_2}$ as $c_1$.
In order to search the lines of marginal stability,
we choose the value of $\zeta_i$ to satisfy the condition (\ref{condition}).
In other words, we move each alignment $u_i(n)~(i=1,2,3)$
to satisfy the condition (\ref{condition}) keeping the value of 
${\rm Im}\ln w_i$ fixed.
We depict in Figure \ref{decays}
the situation where we obtain the BPS states at threshold
for the process (\ref{decay2}).
Then, we can easily read off the condition for the decay (\ref{decay2})
to be possible from Figure \ref{decays} and (\ref{r-param})
\begin{eqnarray}
(a)&:& \zeta_1=0,\nonumber\\
(b)&:& \zeta_1+2\zeta_2=0.
\end{eqnarray}
These expressions are none other than the results obtained in \cite{FiMa}.
\section{Conclusions and discussion}
\hspace{5mm}
In this paper, we have used the idea of \cite{HoVa, HoIqVa} in order to
examine the BPS spectrum of D-branes on the ALE space ${\cal M}$.
We have found that the mirror LG theory can provide a quite useful
framework to study the BPS spectrum.
We have explicitly constructed special Lagrangian submanifolds and 
BPS charge lattice in the mirror manifold ${\widehat {\cal M}}$.
Using these ingredients, we have shown how the lines of marginal stability
can be determined geometrically.
The prescription is much simpler than that of \cite{FiMa}.
Our work provides the framework to see the fate of BPS states in the
whole moduli space of ${\cal M}$.
Although these results are successful,
it would be further desired to clarify the role of BPS algebra found in
\cite{FiMa} in our mirror framework.

In the case of the ALE space ${\cal M}$ with 16 supercharges,
there is much simplification in describing the BPS D-brane moduli space.
In particular, all the lines of marginal stability 
cannot give the decay of BPS states.
This fact is deeply connected to the well-established McKay
correspondence.
It is curious to note that our work on a mirror LG theory has appeared
after the work \cite{FiMa} on a quiver gauge theory.
In the case of Calabi-Yau threefolds, many works have used 
mirror symmetry \cite{BrDoLaRo, DiGo} at first.
This is deeply related to the fact that 
a precise mathematical framework of ``McKay
correspondence'' for Calabi-Yau threefolds
is still out of reach (however, see \cite{DoGrMo} and \cite{Mu}).
Also, it is not obvious how we should define 
a concept of the stability.
Then, one may have a natural question how these two notions are related.
Recently, the study on this direction is quite active \cite{DiDo, GoJa, Do}.
Researches in this direction would provide us with fruitful insights on
the moduli space of D-branes in the near future.

At a different point in the moduli space of ${\cal M}$,
there exists a certain LG description \cite{OoVa, GiKuPe, Le}.
Recently, the modular invariance for the ALE space has been studied
using a free field approximation \cite{EgSu}.
We believe that it would be very interesting 
to understand D-branes in this approach 
in order to fully understand BPS spectrum in the whole moduli space of 
${\cal M}$.

\section*{Acknowledgements}
\hspace{5mm}
M.N. would like to thank K. Hori for his clear lecture 
on mirror symmetry in Summer Institute 2000 at Yamanashi, Japan.
M.N. is grateful to M.R. Douglas for useful communications. The
work of J.H. was supported in part by the Japan Society for the
Promotion of Science under the Postdoctoral Research Program No. 11-09295.
The research of M.N. is supported 
by the JSPS fellowships for Young Scientists. 

\end{document}